\documentstyle[stwol,epsfig]{article}

\def\Journal#1#2#3#4{{#1} {\bf #2}, #3 (#4)}

\def\NPB{{\em Nucl. Phys.} B}
\def\PLB{{\em Phys. Lett.}  B}
\def\PRL{\em Phys. Rev. Lett.}
\def\PRD{{\em Phys. Rev.} D}
\def\ZPC{{\em Z. Phys.} C}
\def\APP{{\em Acta Phys. Pol.} B}
\def\PR{\em Phys. Rep.}
\def\IJM{{\em Int. J. Mod. Phys.} A}

\def\LNC{\em Lett. Nuovo Cimento}
\def\MPL{{\em Mod. Phys. Lett.} A}

\def\be{\begin{equation}}
\def\ee{\end{equation}}
\def\bea{\begin{eqnarray}}
\def\eea{\end{eqnarray}}
\def\lsim{\;\rlap{\lower 3.5 pt \hbox{$\mathchar \sim$}} \raise 1pt
 \hbox {$<$}\;}
\def\gsim{\;\rlap{\lower 3.5 pt \hbox{$\mathchar \sim$}} \raise 1pt
 \hbox {$>$}\;}
\def\n{\hspace*{-2.5mm}}
\def\cl{\mathop{{\mbox{Cl}}_2}\nolimits}

\begin{document}

\title{\vskip-2.cm{\baselineskip14pt
\centerline{\normalsize\hfill MPI/PhT/96--113}
\centerline{\normalsize\hfill hep--ph/9610500}
\centerline{\normalsize\hfill October 1996}
}
\vskip.5cm
THEORETICAL ASPECTS OF HIGGS PHYSICS$^{\mbox{\dag}}$
\vskip22cm{\baselineskip14pt\rm
\begin{flushleft}
$^{\mbox{\dag}}$ To appear in the 
{\it Proceedings of the 28th International Conference on High Energy Physics},
Warsaw,
$\phantom{^{\mbox{\dag}}}$
Poland, 25--31 July 1996, edited by A. Wr\'oblewski.
\end{flushleft}}
\vskip-23cm}

\author{BERND A. KNIEHL}

\address{Max-Planck-Institut f\"ur Physik (Werner-Heisenberg-Institut),\\
F\"ohringer Ring 6, 80805 Munich, Germany\\
and\\
Institut f\"ur Theoretische Physik, Ludwig-Maximilians-Universit\"at,\\
Theresienstra\ss e~37, 80333 Munich, Germany}

\twocolumn[\maketitle\abstracts{
We review the present status of Higgs physics within the standard model and its 
extensions.
First, we briefly summarize the current experimental exclusion limits from the
direct searches with LEP1 and the Tevatron, and assess the discovery
potential of LEP2.
Then, we report the mass bounds resulting from global fits to the latest
electroweak precision data, and compile various theoretical principles which
lead to restrictions---or even determinations---of the Higgs-boson masses.
Perturbative upper bounds are discussed in some detail.
Finally, we survey the recent progress in the computation of higher-order
radiative corrections to the $b\bar b$ decay width of the standard-model Higgs
boson.}]

\section{Introduction}

The SU(2)$_I\times$U(1)$_Y$ structure of the electroweak interactions has been
consolidated by an enormous wealth of experimental data during the past three
decades.
The canonical way to generate masses for the fermions and intermediate bosons
without violating this gauge symmetry in the Lagrangian is by the Higgs
mechanism of spontaneous symmetry breaking.
In the minimal standard model (SM), this is achieved by introducing one
complex SU(2)$_I$-doublet scalar field $\Phi$ with $Y=1$.
The three massless Goldstone bosons which emerge via the electroweak symmetry
breaking are eaten up to become the longitudinal degrees of freedom of the
$W^\pm$ and $Z$ bosons, {\em i.e.}, to generate their masses, while one 
CP-even Higgs scalar boson $H$ remains in the physical spectrum.
The Higgs potential contains one mass and one self-coupling.
Since the vacuum expectation value (VEV) is fixed by the relation $v=2M_W/g$,
where $g$ is the SU(2)$_I$ gauge coupling, there remains one free parameter in
the Higgs sector, namely, $M_H$.

A phenomenologically interesting extension of the SM Higgs sector that keeps 
the electroweak $\rho$ parameter at unity in the Born approximation, is
obtained by adding a second complex SU(2)$_I$-doublet scalar field $\Phi_2$
with $Y=-1$.
This leads to the two-Higgs-doublet model (2HDM).
After the three Goldstone bosons have been eliminated, there remain five
physical Higgs scalars: the neutral CP-even $h^0$ and $H^0$ bosons, the
neutral CP-odd $A^0$ boson, and the charged $H^\pm$ pair.
The most general CP-conserving Higgs potential for the 2HDM contains three 
masses and four self-couplings.
Subtracting the $M_W$ constraint on the two VEV's $v_1$ and $v_2$, one is left
with six free parameters, which are usually taken to be $m_{h^0}$, $m_{H^0}$,
$m_{A^0}$, $m_{H^\pm}$, $\tan\beta=v_2/v_1$, and the weak mixing angle
$\alpha$ which relates the weak and mass eigenstates of $h^0$ and $H^0$.
According to the Glashow-Weinberg theorem,\cite{gla} flavour-changing neutral 
currents may be avoided if all fermions with the same electric charge couple to
the same Higgs doublet.
In the 2HDM of type~II, the up-type/down-type fermions couple to the Higgs 
doublet with $Y=\pm1$.

The Higgs sector of the minimal supersymmetric extension of the SM (MSSM) 
consists of a 2HDM of type~II.
Since the Higgs self-couplings are then determined by the gauge couplings, 
there are only two free parameters in the MSSM Higgs sector, which are 
conveniently chosen to be $m_{A^0}$ and $\tan\beta$.
However, a large number of new masses, couplings, and mixing angles connected
with the supersymmetric partners enter via loop effects.
In the supergravity-inspired MSSM, all these degrees of freedom may be related
to just five parameters at the GUT scale:
the Higgs mixing $\mu$, the sfermion mass $m_0$, the gaugino mass $m_{1/2}$, 
the trilinear sfermion-Higgs coupling $A$, and the bilinear Higgs coupling 
$B$.
After the renormalization-group (RG) evolution down to the electroweak scale,
the heavy sfermions are approximately degenerate, with mass $M_S$.
The following three scenarios are often considered in the literature:
$(i)$ no mixing: $A=0$, $|\mu|\ll M_S$;
$(ii)$ typical mixing: $A=-\mu=M_S$; and
$(iii)$ maximal mixing: $A=\sqrt6M_S$, $|\mu|\ll M_S$.

\section{Direct Higgs searches}

\subsection{LEP1 and Tevatron}

\begin{figure}[ht]
\center
\epsfig{figure=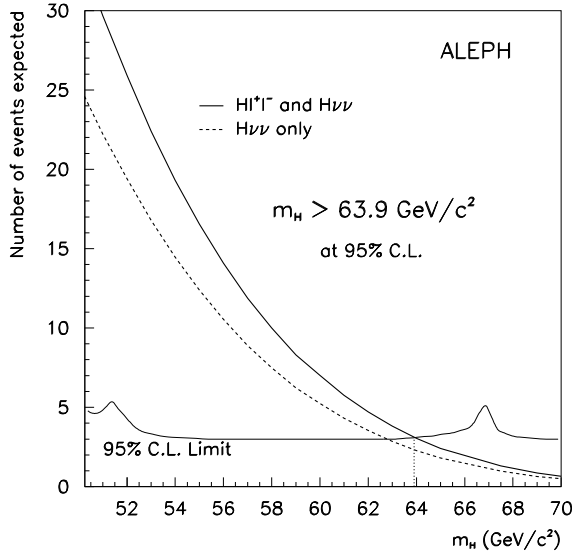,width=7.5cm}
\caption{$M_H$ dependence of the number of Higgs signal events expected by
ALEPH.}
\label{fig:aleph}
\end{figure}

The SM Higgs boson was searched at LEP1 via Bjorken's process
$e^+e^-\to Hf\bar f$ by looking for a pair of acoplanar jets together with
missing energy in the $H\nu\bar\nu$ channel, or for a hadronic event with an
energetic lepton pair in the $Hl^+l^-$ ($l=e,\mu$) channel.
The extraction by ALEPH of the 95\% CL lower bound on $M_H$ is illustrated in
Fig.~\ref{fig:aleph}.
The present LEP1 bounds~\cite{bus,mar} are summarized in Table~\ref{tab:mh}.

\begin{table}\begin{center}\caption{95\% CL $M_H$ lower bounds (in GeV) from
LEP1.}
\label{tab:mh}
\vspace{0.4cm}
\begin{tabular}{|c|c|c|c|} \hline
Collab. & Years & $10^6$ $q\bar q$ & $M_H^{\rm min}$ \\ \hline
ALEPH & 89--95 & 4.5 & 63.9 \\
DELPHI & 91, 92, 94, 95 & 3.1 & 58.3 \\
L3 & 91--94 & 3.1 & 60.2 \\
OPAL & 89--95 & 4.4 & 59.6 \\ \hline
\end{tabular}
\end{center}
\end{table}

\begin{figure}[ht]
\center
\epsfig{figure=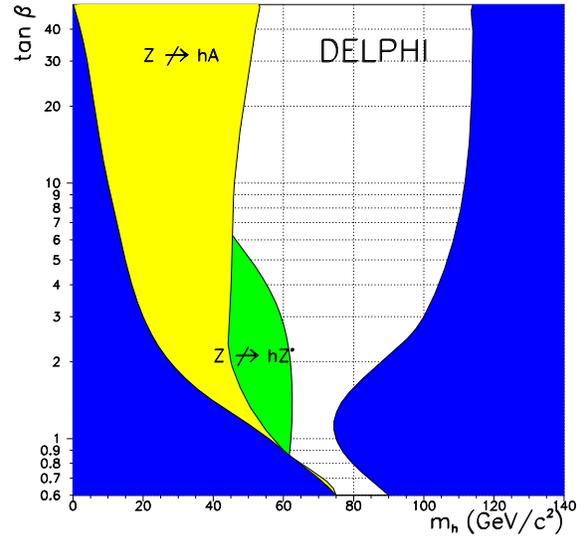,width=7.5cm}
\caption{Region in the $(m_{h^0},\tan\beta)$ plane excluded at 95\% CL by 
DELPHI. The dark areas are forbidden by theory.}
\label{fig:delphi}
\end{figure}

The $h^0$ and $A^0$ bosons of the 2HDM or MSSM could have been produced at
LEP1 via Higgs-strahlung $e^+e^-\to h^0f\bar f$, where the $h^0$ boson is
radiated off the resonant $Z$ boson, associated production $e^+e^-\to h^0A^0$,
and the Yukawa process $e^+e^-\to h^0/A^0f\bar f$ ($f=b,\tau$), where the
$h^0$ and $A^0$ bosons couple to the fermion line.
The $h^0Z^*$ and $h^0A^0$ channels are complementary, in the sense that their
cross sections are proportional to $\sin^2(\beta-\alpha)$ and
$\cos^2(\beta-\alpha)$, respectively.
In the framework of the MSSM with $M_S=1$~TeV and no squark mixing, DELPHI has
excluded the shaded regions in the $(m_{h^0},\tan\beta)$ plane of
Fig.~\ref{fig:delphi}.
The current 95\% CL lower bounds on $m_{h^0}$ and $m_{A^0}$ in the MSSM
obtained by the four LEP1 experiments~\cite{mar,jan} are listed in 
Table~\ref{tab:mssm}.

\begin{table}\begin{center}\caption{95\% CL lower bounds on $m_{h^0}$,
$m_{A^0}$, and $m_{H^\pm}$ (in GeV) from LEP1.}
\label{tab:mssm}
\vspace{0.4cm}
\begin{tabular}{|c|c|c|c|} \hline
Collab. & $m_{h^0}^{\rm min}$ & $m_{A^0}^{\rm min}$ & $m_{H^\pm}^{\rm min}$ \\
\hline
ALEPH & 45.5 & 45 & 41.7 \\
DELPHI & 45.4 & 45.2 & -- \\
L3 & 42 & 23 & 42.8 \\
OPAL & 44.3 & 23.5 & 44.1 \\ \hline
\end{tabular}
\end{center}
\end{table}

In the more general context of the 2HDM, the $h^0$ ($A^0$) boson could have
been missed in the $h^0Z^*$ and $h^0A^0$ channels, if
$\sin^2(\beta-\alpha)\approx0$ (1) and (or) $m_{h^0}+m_{A^0}>M_Z$.
Then, the Yukawa process would be dominant.\cite{kal}
The MSSM scenario with $m_{A^0}\ll M_Z$ and $\tan\beta\gg1$ was considered
particularly interesting,\cite{wel} since it could partially explain the
$R_b$ anomaly, which was then still existing. 
ALEPH~\cite{yuk} searched for the $h^0$ and $A^0$ bosons of the 2HDM using the
Yukawa process with $b\bar bb\bar b$, $\tau^+\tau^-b\bar b$,
$\tau^+\tau^-\tau^+\tau^-$, and $\tau^+\tau^-X$ final states, where $X$ is a 
system with low charged multiplicity.
The $b\bar bb\bar b$ channel turned out to be less efficient in ruling out the
small-$m_{A^0}$ large-$\tan\beta$ scenario than anticipated.\cite{wel}
It is amusing to observe that, in the 2HDM with $\alpha\approx\beta$, the most
stringent $m_{h^0}$ lower bound presently comes from HERA.\cite{baw}
The future E821 experiment at BNL is expected to decrease the experimental 
error on $(g-2)_\mu$ by a factor of twenty or more,\cite{rob} and will thus
lead to interesting $m_{h^0}$ and $m_{A^0}$ lower bounds via loop
effects.\cite{kra}

\begin{figure}[ht]
\center
\epsfig{figure=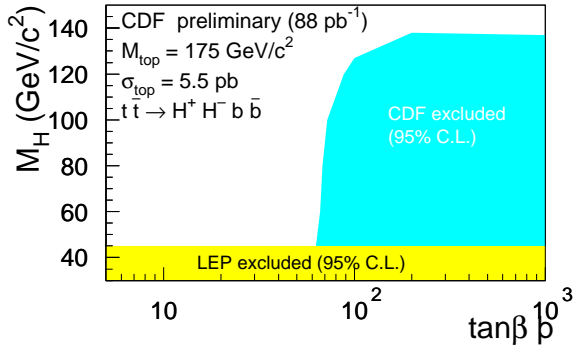,width=7.5cm}
\caption{Region in the $(\tan\beta,m_{H^\pm})$ plane excluded at 95\% CL by
CDF.}
\label{fig:cdf}
\end{figure}

Searching for the $H^\pm$ bosons of the 2HDM via pair production
$e^+e^-\to H^+H^-$, the LEP1 experiments~\cite{dec} could almost exclude the
entire $m_{H^\pm}$ range allowed by kinematics (see Table~\ref{tab:mssm}).
CDF~\cite{abe} has looked for $H^\pm$ bosons in the decay products of
$t\bar t$ pairs, and has greatly increased the LEP1 lower bound on $m_{H^\pm}$
for $\tan\beta\gsim60$ (see Fig.~\ref{fig:cdf}).
If $\tan\beta\gg1$, then the cascade $t\to H^+b\to\tau^+\nu_\tau b$ is more 
likely to happen than $t\to W^+b\to\tau^+\nu_\tau b$ and leads to a larger
missing transverse energy.\cite{abe}

\subsection{LEP2}

\begin{figure}[ht]
\center
\epsfig{figure=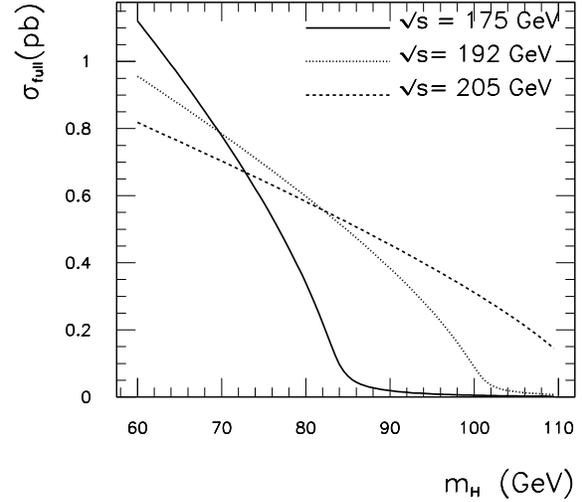,width=7.5cm}
\caption{Cross section of Higgs-strahlung at LEP2 as a function of $M_H$.}
\label{fig:gross}
\end{figure}

The theoretical and experimental aspects of Higgs phenomenology at LEP2 have 
recently been sum\-mar\-ized in a comprehensive report.\cite{car}
Higgs-strahlung is the dominant production mechanism of the SM Higgs boson at 
LEP2.
In Fig.~\ref{fig:gross}, its cross section is shown as a function of $M_H$ for
three values of CM energy.\cite{gro}
Electroweak radiative corrections~\cite{fle} are included here.

\begin{figure}[ht]
\center
\epsfig{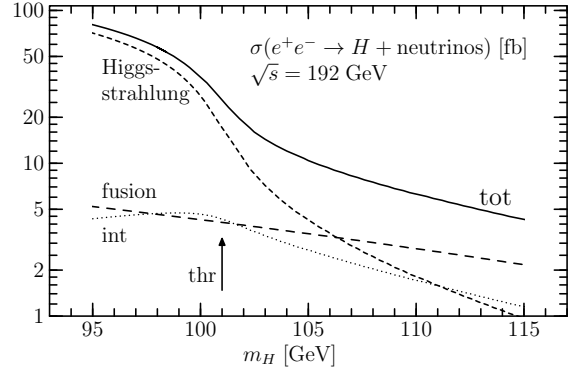}
\caption{$H\nu_e\bar\nu_e$ production via Higgs-strahlung, $W^+W^-$ fusion,
and their interference in the threshold region at LEP2.}
\label{fig:kilian}
\end{figure}

The $H\nu_e\bar\nu_e$ and $He^+e^-$ final states are also produced through
$W^+W^-$ and $ZZ$ fusion, respectively.
In the threshold region, the signal cross section is appreciably increased
by coherently including these processes.\cite{kil}
This is illustrated for $H\nu_e\bar\nu_e$ production in Fig.~\ref{fig:kilian}.

\begin{figure}[ht]
\center
\epsfig{figure=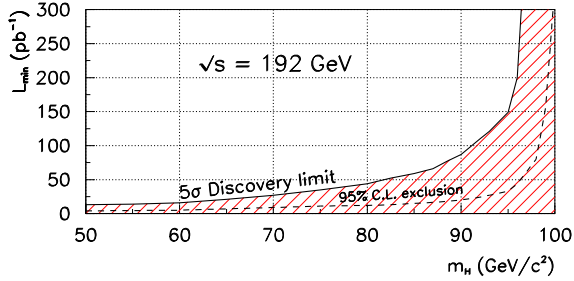,width=7.5cm}
\caption{LEP2 exclusion and discovery limits for the SM Higgs boson.}
\label{fig:smh}
\end{figure}

The minimum luminosity $L_{\rm min}$ needed per experiment for a combined
$5\sigma$ discovery or a 95\% CL exclusion of the SM Higgs boson under
realistic LEP2 conditions~\cite{car} is shown as a function of $M_H$ in
Fig.~\ref{fig:smh}.
We see that, at $\sqrt s=192$~GeV, $L_{\rm min}=150$~pb$^{-1}$ is sufficient 
to discover the SM Higgs boson with $M_H\lsim95$~GeV.
On the other hand, a 95~GeV Higgs boson can be excluded at 95\% CL with
$L_{\rm min}$ as low as 33~pb$^{-1}$, while, with $L_{\rm min}=150$~pb$^{-1}$,
the $M_H$ range way up to 98~GeV may be excluded.

\begin{figure}[ht]
\center
\epsfig{figure=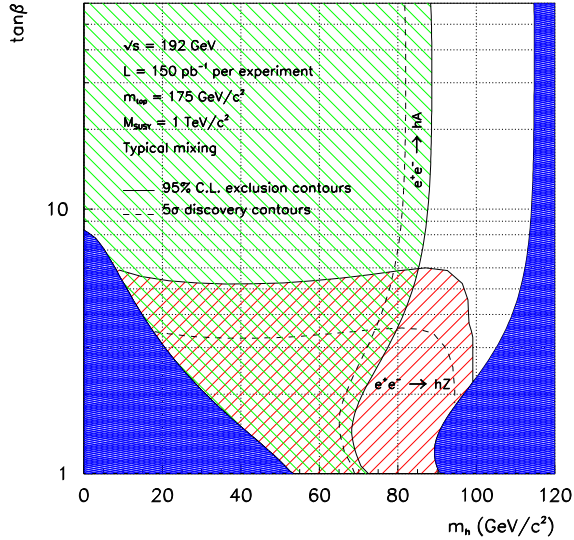,width=7.5cm}
\caption{LEP2 exclusion and discovery limits in the $(m_{h^0},\tan\beta)$
plane.}
\label{fig:mssmh}
\end{figure}

The $5\sigma$ discovery and 95\% CL exclusion limits in the
$(m_{h^0},\tan\beta)$ plane of the MSSM that may be reached at LEP2 with a
luminosity of 150~pb$^{-1}$ per experiment~\cite{car} may seen from
Fig.~\ref{fig:mssmh}.
Here, $M_t=175$~GeV, $M_S=1$~TeV, and typical squark mixing as described in 
the Introduction are assumed.
Comparing Figs.~\ref{fig:delphi} and \ref{fig:mssmh}, we observe that LEP2
will be able to close the low-$\tan\beta$ region which was not covered by LEP1.

\section{Higgs mass bounds}

\subsection{Global fits}

\begin{figure}[ht]
\center
\epsfig{figure=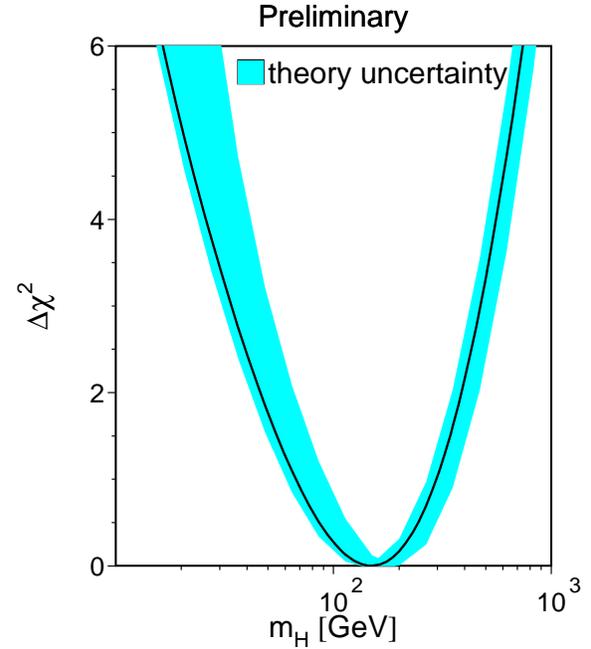,width=7.5cm}
\caption{$M_H$ dependence of $\Delta\chi^2=\chi^2-\chi_{\rm min}^2$ resulting
from a global fit to the latest experimental data.}
\label{fig:ewwg}
\end{figure}

Experimental precision tests of the standard electroweak theory are
sensitive to the Higgs boson via quantum corrections.
A global fit~\cite{alc} to the $e^+e^-$ data on the $Z$-boson observables from
ALEPH, DELPHI, L3, OPAL, and SLD,
the $p\bar p$ data on $M_W$ from CDF, D0, and UA2,
the $\nu N$ data on $\sin^2\theta_w$ from CCFR, CDHS, and CHARM, and
the Tevatron data on $M_t$ from CDF and D0 which were presented during this
conference yields $M_H=149{+148\atop-82}$GeV with
$\chi_{\rm min}^2/\mbox{d.o.f.}=19/14\approx1.36$.
The resulting $M_H$ distribution of $\Delta\chi^2=\chi^2-\chi_{\rm min}^2$ is
shown in Fig.~\ref{fig:ewwg}, where the shaded band represents the
estimated~\cite{alc} theoretical error due to missing higher-order 
corrections.
From Fig.~\ref{fig:ewwg}, one may read off an 95\% CL upper bound on $M_H$ at
550~GeV.\cite{alc}
The electroweak precision data have reached a quality which even allows for 
global fits within the MSSM.\cite{boe}

\subsection{Theoretical principles}

There have been various attempts to bound or even determine $M_H$ from first
principles.

\subsubsection{Naturalness}

By requiring that the Higgs-boson mass counterterm, {\em i.e.}, the difference
between the bare and renormalized masses, be devoid of quadratic ultraviolet 
divergences, Veltman derived the mass relation~\cite{vel}
\be
\sum_fN_fM_f^2=\frac{3}{2}M_W^2+\frac{3}{4}M_Z^2+\frac{3}{4}M_H^2,
\ee
where $N_f=1$ (3) for leptons (quarks).
Na\"\i vely inserting the known pole masses, this leads to $M_H\approx320$~GeV.
Since this prediction is based on fine-tuning, it is nowadays considered
unnatural.

\subsubsection{Noncommutative geometry}

It is possible~\cite{hau} to construct the SM on the basis of the graded Lie
algebra SU(2$|$1), which contains the SU(2)$_I\times$U(1)$_Y$ Lie algebra in
its even part.
The essential ingredient is an algebraic superconnection which incorporates 
both the gauge and Higgs fields and whose curvature automatically generates 
the spontaneously broken realization of the SM.
This leads to a geometrical interpretation of the Higgs mechanism.
As a by-product, one obtains~\cite{hau} the mass relation $M_H=\sqrt2M_W$.
Taking this as an initial condition at the GUT scale and performing a one-loop
RG analysis, one finds~\cite{oku} the physical Higgs-boson mass to be
$M_H=164$~GeV if $M_t=175$~GeV.

\subsubsection{Triviality and vacuum stability}

Triviality~\cite{cab} and vacuum stability~\cite{lin} are features connected
with the RG-improved effective potential
$V_{\rm eff}=-m^2(\mu)|\Phi|^2+\lambda(\mu)|\Phi|^4/2$.
The dependence of the quartic self-coupling $\lambda(\mu)$ on the
renormalization scale $\mu$ is determined by the RG equations.
Roughly speaking, the triviality upper bound (vacuum-stability lower bound) on
$M_H$ follows from the requirement that $\lambda(\mu)$ stay finite (positive)
for all $\mu<\Lambda$, where $\Lambda$ is the cutoff beyond which new physics
operates.
Assuming the SM to be valid up to the GUT scale $\Lambda=10^{16}$~GeV, one 
thus obtains~\cite{car} 130~GeV${}\lsim M_H\lsim{}$180~GeV.
In turn, should the SM Higgs boson be discovered at LEP2, new physics is 
expected to appear below $\Lambda=10$~TeV.

\subsubsection{Metastability}

Depending on $M_H$ and $M_t$, $V_{\rm eff}$ at finite (or zero) temperature
may have a deep stable minimum at
$\langle|\Phi_{\rm min}|\rangle\gg G_F^{-1/2}$,
{\em i.e.}, the physical electroweak minimum may just be metastable.
Then, an absolute lower bound on $M_H$ follows from the condition that the
probability, normalized w.r.t. the expansion rate of the Universe, for the
decay of the metastable vacuum by thermal fluctuations (or quantum tunneling)
be negligibly small.\cite{and}
The $M_H$ bounds thus obtained are somewhat below the usual stability
bounds.\cite{lin}

\subsubsection{Multiple point criticality principle}

It has been argued~\cite{fro} that a mild form of locality breaking in quantum 
gravity due to baby universes, which are expected to render coupling constants 
dynamical, leads to the realization of the so-called multiple-point
criticality principle in Nature.
According to this principle, Nature should choose coupling constants such that
the vacuum can exist in degenerate phases, {\em i.e.},
$V_{\rm eff}(\Phi_{\rm min,1})=V_{\rm eff}(\Phi_{\rm min,2})$.
In order that the dynamical mechanism be relevant, these authors~\cite{fro}
also require a strong first-order phase transition between the two vacua,
implemented by
$\langle|\Phi_{\rm min,2}|\rangle\approx M_{\rm Planck}=2\times10^{19}$~GeV.
Via the usual RG analysis, these two assumptions then lead to a simultaneous
prediction of $M_t$ and $M_H$, namely,\cite{fro} $M_t=(173\pm4)$~GeV and
$M_H=(135\pm9)$~GeV.

\subsubsection{Perturbation-theory breakdown}

An attractive way of constraining $M_H$ from above is to require that the
Higgs sector be weakly interacting, so that perturbation theory is
meaningful.\cite{mve}
The resulting bounds depend somewhat on the considered process and the precise
definition of perturbation-theory breakdown, but they are independent of
assumptions concerning the scale $\Lambda$ of new physics.
At one and two loops, the leading high-$M_H$ corrections to physical
observables related to Higgs-boson production or decay are of
${\cal O}(G_FM_H^2)$ and ${\cal O}(G_F^2M_H^4)$, respectively.
For $M_H$ increasing, the ${\cal O}(G_F^2M_H^4)$ corrections eventually exceed
the ${\cal O}(G_FM_H^2)$ ones in size, so that the perturbative expansions in
$G_FM_H^2$ cease to usefully converge.
The values $M_H^{\rm max}$ where this happens may be used to define a
perturbative upper bound on $M_H$.

\begin{figure}[ht]
\center
\epsfig{figure=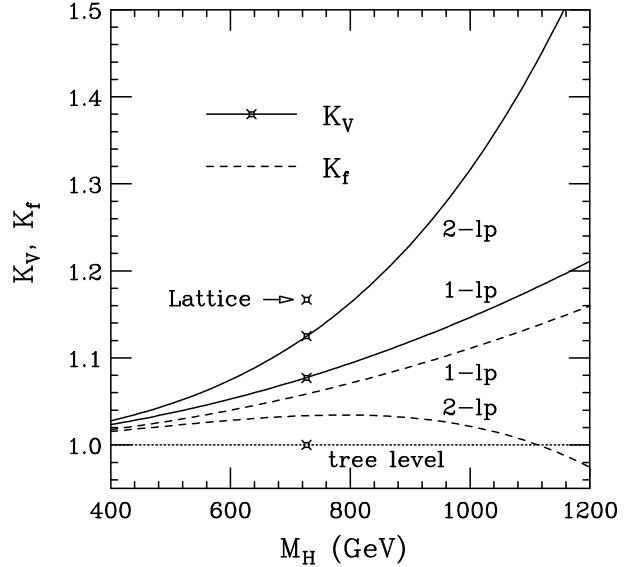,width=7.5cm,angle=90}
\caption{$K_V$ and $K_f$ to ${\cal O}(G_FM_H^2)$ and ${\cal O}(G_F^2M_H^4)$ as
functions of $M_H$.
The crosses indicate the tree-level, one-loop, two-loop, and nonperturbative
values of $K_V$ at $M_H=727$~GeV.}
\label{fig:frink}
\end{figure}

As first examples, the Higgs decays to pairs of fermions~\cite{dur} and
intermediate bosons~\cite{ghi} have recently been studied through
${\cal O}(G_F^2M_H^4)$.
This task may be greatly simplified in the limit of interest, $M_H\gg2M_Z$,
through the use of the Goldstone-boson equivalence theorem.\cite{cor}
This theorem states that the leading high-$M_H$ electroweak contribution to a
Feynman diagram may be calculated by replacing the intermediate bosons
$W^\pm$ and $Z$ with the respective would-be Goldstone bosons $w^\pm$ and $z$
of the symmetry-breaking sector.
In this limit, the gauge and Yukawa couplings may be neglected against the
Higgs self-coupling.
By the same token, the Goldstone bosons may be taken to be massless, and the
fermion loops may be omitted.
The resulting correction factor $K_f$ for $\Gamma\left(H\to f\bar f\,\right)$
is independent of the fermion flavour $f$.
Similarly, $\Gamma(H\to W^+W^-)$ and $\Gamma(H\to ZZ)$ receive the same 
correction factor $K_V$.
In the on-mass-shell renormalization scheme, the results read~\cite{dur,ghi}
\bea
\label{gbet}
K_f&\n=\n&1+\hat\lambda\left(13-2\pi\sqrt3\right)
+\hat\lambda^2\left[12-169\pi\sqrt3
\right.\nonumber\\
&\n\n&{}+170\zeta(2)-252\zeta(3)+12\left(13\pi+19\sqrt3\right)
\nonumber\\
&\n\n&{}\times\left.\cl\left(\frac{\pi}{3}\right)\right]
\nonumber\\
&\n\approx\n&1+11.1\%\left(\frac{M_H}{\mbox{TeV}}\right)^2
-8.9\%\left(\frac{M_H}{\mbox{TeV}}\right)^4,
\nonumber\\
K_V&\n=\n&1+\hat\lambda\left(19-6\pi\sqrt3-10\zeta(2)\right)
+62.0\,\hat\lambda^2
\nonumber\\
&\n\approx\n&1+14.6\%\left(\frac{M_H}{\mbox{TeV}}\right)^2
+16.9\%\left(\frac{M_H}{\mbox{TeV}}\right)^4\!\!,
\eea
where $\hat\lambda=\left(G_FM_H^2/16\pi^2\sqrt2\,\right)$,
$\zeta$ is Riemann's zeta function, and $\cl{}$ is Clausen's integral.
The ${\cal O}(G_FM_H^2)$ terms in Eq.~(\ref{gbet}) have been known for a long
time.\cite{vlt}
$K_f$ and $K_V$ are displayed as functions of $M_H$ in Fig.~\ref{fig:frink},
from where we may read off the $M_H^{\rm max}$ values 1114~GeV and 930~GeV,
respectively.
The nonperturbative value of $K_V$ at $M_H=727$~GeV may be extracted from a
recent lattice simulation of elastic $\pi\pi$ scattering in the framework of
the four-dimensional O(4)-symmetric nonlinear $\sigma$ model in the broken
phase, where the $\sigma$ resonance was observed.\cite{goe}

\begin{figure}[ht]
\center
\epsfig{figure=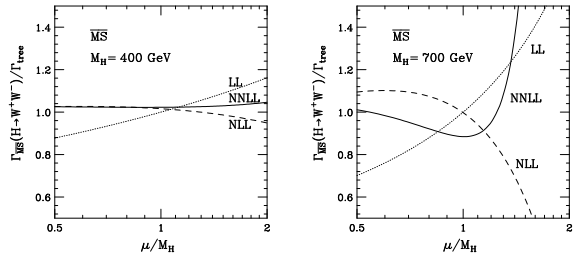,width=7.5cm}
\caption{$\mu$ dependence of $K_V$ in the $\overline{\mbox{MS}}$ scheme at
$M_H=400$ and 700~GeV.}
\label{fig:kurt}
\end{figure}

The study of the $\mu$ dependence of Higgs-boson observables in the
$\overline{\mbox{MS}}$ renormalization scheme provides another aspect of
perturbation-theory breakdown in the Higgs sector.
If perturbation theory is valid, the $\mu$ dependence should be reduced as
higher-order corrections are included.
While this empirical rule is satisfied for $\Gamma(H\to W^+W^-)$ at
$M_H=400$~GeV, it is clearly violated at $M_H=700$~GeV,\cite{kurt} as may be
seen from Fig.~\ref{fig:kurt}.

\section{Radiative corrections to $\Gamma\left(H\to b\bar b\right)$}

The SM Higgs boson with intermediate mass $M_H\lsim2M_W$ decays dominantly to
$b\bar b$ pairs.
The radiative corrections to the partial width of this decay involve two very
different scales, namely, $M_H$ and $M_t$.
It is therefore convenient to treat this process in the framework of a $n_f=5$ 
effective Yukawa Lagrangian, {\em i.e.}, to integrate out the top quark.
This leads to a RG-improved formulation, which provides a natural separation
of the $n_f=5$ QCD corrections at scale $\mu=M_H$ and the top-quark-induced
$n_f=6$ corrections at scale $\mu=M_t$.
One thus obtains~\cite{cks} the following structure:
\bea
\Gamma_{b\bar b}&\n=\n&\Gamma_{b\bar b}^{\rm Born}
\left[\left(1+\Delta_b^{\rm QED}\right)
\left(1+\left.\Delta_b^{\rm weak}\right|_{x_t=0}\right)\right.
\nonumber\\
&\n\n&{}\times\left.\left(1+\Delta_b^{\rm QCD}\right)
\left(1+\Delta_b^t\right)+\Xi_b^{\rm QCD}\Xi_b^t\right].\quad
\eea
If the Born formula
\be
\Gamma_{b\bar b}^{\rm Born}=\frac{3G_FM_Hm_b^2}{4\pi\sqrt2}
\left(1-\frac{4m_b^2}{M_H^2}\right)^{3/2}
\ee
is written with the QED and QCD $\overline{\mbox{MS}}$ mass evaluated with
$n_f=5$ quark flavours at scale $\mu=M_H$, $m_b^{(5)}(M_H)$, then
$\Delta_b^{\rm QED}$ and $\Delta_b^{\rm QCD}$ are finite for $m_b=0$ and
read~\cite{bra}
\bea
\Delta_b^{\rm QED}&\n=\n&\frac{17}{36}\,\frac{\alpha}{\pi},
\\
\Delta_b^{\rm QCD}&\n=\n&\frac{17}{3}a_5
+a_5^2\left(\frac{8851}{144}-\frac{47}{6}\zeta(2)-\frac{97}{6}\zeta(3)\right)
\nonumber\\
&\n\n&{}+a_5^3\left(\frac{34873057}{46656}-\frac{10225}{54}\zeta(2)
\right.\nonumber\\
&\n\n&{}-\left.
\frac{80095}{216}\zeta(3)-\frac{25}{6}\zeta(4)+\frac{1945}{36}\zeta(5)\right)
\nonumber\\
&\n=\n&5.66667\,a_5+29.14671\,a_5^2+41.75760\,a_5^3,
\nonumber
\eea
where $a_5=\alpha_s^{(5)}(M_H)/\pi$.
By means of scale optimization according to the principles of fastest apparent 
convergence (FAC) or minimal sensitivity (PMS), the coefficient of $a_5^4$ may
be estimated~\cite{sir} to be $-981$.
In the limit $M_H\ll2M_W$, the weak correction, with the ${\cal O}(G_FM_t^2)$
term stripped off, reads~\cite{kni}
\bea
\left.\Delta_b^{\rm weak}\right|_{x_t=0}&\n=\n&\frac{G_FM_Z^2}{8\pi^2\sqrt2}
\left(\frac{1}{6}-\frac{7}{3}c_w^2-\frac{16}{3}c_w^4
\right.\nonumber\\
&\n\n&{}+\left.
3\frac{c_w^2}{s_w^2}\ln c_w^2\right),
\eea
where $c_w^2=1-s_w^2=M_W^2/M_Z^2$.
The leading contributions due to the $b\bar b$ and $b\bar bg$ cuts of the
double-triangle diagrams where the top quark circulates in one of the
triangles are contained in~\cite{cks,lar}
\be
\Xi_b^{\rm QCD}=a_5\left(-\frac{76}{3}+8\zeta(2)
-\frac{4}{3}\ln^2\frac{m_b^2}{M_H^2}\right).
\ee
The would-be mass singularity proportional to $\ln^2(m_b^2/M_H^2)$ cancels if
the $b\bar b(g)$ and $gg$ decay channels are combined.\cite{lar}
The leading top-quark-induced corrections are concentrated in~\cite{cks}
\bea
\Delta_b^t&\n=\n&
a_6^2\left(\frac{5}{9}+\frac{2}{3}L\right)
+x_t\left[1+a_6\left(-\frac{4}{3}-4\zeta(2)\right)
\right.\nonumber\\
&\n\n&{}+\left.
a_6^2\left(-59.16260+\frac{2}{3}L\right)\right],
\nonumber\\
\Xi_b^t&\n=\n&a_6\left(-\frac{1}{12}-\frac{1}{12}x_t\right),
\end{eqnarray}
where $\mu_t=m_t^{(6)}(\mu_t)$, $a_6=\alpha_s^{(6)}(\mu_t)/\pi$,
$L=\ln(\mu_t^2/M_H^2)$, and $x_t=\left(G_F\mu_t^2/8\pi^2\sqrt2\,\right)$.
The $L$-dependent terms may be resummed by exploiting the RG-invariance of the 
energy-momentum tensor.\cite{cks}

The QCD correction~\cite{ina} to $\Gamma(H\to gg)$ includes a contribution due
to the
$b\bar bg$ final state, which may also be interpreted~\cite{zer} as a
${\cal O}(\alpha_s^3M_H^2/m_b^2)$ correction to
$\Gamma\left(H\to b\bar b\right)$.

\section*{Acknowledgments}

The author is indebted to Ralf Hempfling for beneficial discussions.

\end{document}